\documentclass[12pt]{article}
\usepackage{amsmath,amsfonts,amsthm,amssymb,slashed,graphicx,mathtools,bbm}
\usepackage{xcolor}
\usepackage{array,booktabs}
\usepackage{hyperref}
\usepackage[square, comma,numbers,sort&compress]{natbib}

\numberwithin{equation}{section}
\newcommand{\beq}{\begin{equation}}
\newcommand{\eeq}{\end{equation}}

\newcommand{\address}[1]{\vbox{\center\em#1}}
\renewcommand{\title}[1]{\vbox{\center\LARGE{#1}}\vspace{5mm}}

\DeclareMathOperator{\Tr}{Tr}

\makeatletter
\newcommand*{\letterdef@}{}
\newcommand*{\letterdef}[3]{%
  \def\letterdef@##1{\expandafter\newcommand\csname #1\endcsname{#2{##1}}}%
  \@tfor\@tempa :=#3\do{\expandafter\letterdef@\expandafter{\@tempa}}}
\makeatother
\letterdef{b#1} {\mathbb} {CHPRZ}    
\letterdef{c#1} {\mathcal}{ABCDEFGHIJKLMNOPQRSTUVWXYZ} 
\letterdef{rm#1}{\mathrm} {dDF} 

\newcommand\para{\paragraph{}}

\begin{document}

\begin{titlepage}
\hfill{preprint SISSA 14/2018/FISI}
\\
\vspace{15mm}
\center{{\LARGE $T\bar T$-deformations in closed form}}

\vspace{10mm}

\begin{center}
\renewcommand{\thefootnote}{$\alph{footnote}$}
Giulio Bonelli, Nima Doroud and Mengqi Zhu

\vspace{10mm}

\address{International School for Advanced Studies (SISSA) \\
Via Bonomea 265, 34136 Trieste, Italy\\ \&\\ INFN Sezione di Trieste }

\renewcommand{\thefootnote}{\arabic{footnote}}
\setcounter{footnote}{0}

\end{center}

\vspace{25mm}

\abstract{
\medskip\medskip
\normalsize{
We consider the problem of exact integration of the $T\bar T$-deformation of two dimensional quantum field theories, as well as some higher dimensional extensions in the form of $\det T$-deformations. When the action can be shown to only depend algebraically on the background metric the solution of the deformation equation on the Lagrangian can be given in closed form in terms of solutions of the (extended) Burgers' equation. We present such examples in two and higher dimensions.
}
}

\vspace{10mm}

\noindent
\vfill

\end{titlepage}

\tableofcontents


\section{Introduction}

Exact quantification of how Quantum Field Theories react as we vary the coupling constants or dynamical scales is a crucial issue in modern theoretical physics \cite{Wilson:1993dy,Polchinski:1983gv}. Cases where such deformations can be integrated exactly and in closed form are extremely rare and often enjoy supersymmetry. In this framework, QFTs in two dimensions reveal to be special since there exists examples of non-supersymmetric interacting theories which are integrable and whose renormalisation group flow can be determined explicitly \cite{AAR}.

\para
A generic QFT admits deformations by operators which instigate a flow as we probe the dynamics at different scales. As one is often interested in the dynamics at low energy scales, deformations which drive the flow at lower energies are considered \emph{relevant} whilst deformations that dominate the flow as we probe the dynamics at higher energy scales are considered \emph{irrelevant}. The latter flow is much harder to study as it generally involves reintroducing the high energy degrees of freedom that have been integrated out. Nonetheless there are examples for which the flow can be determined, notably the deformation of any local relativistic QFT in two spacetime dimensions by the irrelevant $T\bar{T}$ operator first engineered by Zamolodchikov in \cite{Zamolodchikov:2004ce} (See also \cite{Fateev:1998xb}).

\para
Although the construction of the $T\bar{T}$-operator in \cite{Zamolodchikov:2004ce} holds true for general $D=2$ QFTs, subsequent studies mostly focused on integrable quantum field theories (IQFT) \cite{Delfino:2004vc,Caselle:2013dra,Dubovsky:2012wk,Dubovsky:2013gi}. This is due to the fact that the $T\bar{T}$-deformation of IQFTs preserves the integrable structure (See for instance \cite{Mussardo:1999aj}) providing a better handle on the dynamics at high energies. This was recently proved and generalised to an infinite class of irrelevant deformations of IQFTs in two dimensions by Smirnov and Zamolodchikov in \cite{Smirnov:2016lqw}. Their results along with \cite{Cavaglia:2016oda} sparked a renewed interest in irrelevant deformations of quantum field theories with various generalisations proposed in \cite{Guica:2017lia,Cardy:2018sdv,Baggio:2018gct} and applications to holography studied in \cite{McGough:2016lol,Asrat:2017tzd,Giveon:2017nie,Giveon:2017myj,vanLeuven:2018pwv,Shyam:2017znq,Giribet:2017imm,Bzowski:2018pcy}. Some implications of this irrelevant deformation for the UV theory were considered in \cite{Cottrell:2018skz,Aharony:2018vux,Dubovsky:2017cnj}, and in \cite{Bernard:2015bba} a hydrodynamical approach was considered.

\para
The aim of this paper is to study the $T\bar T$-deformation of QFTs in $D=2$, and its extensions to higher dimensions, \emph{both for conformal and for massive theories}. More specifically, the flow equation induced by the $T\bar T$-deformation can be reformulated as a functional equation. Under certain conditions the functional equation reduces to a simple PDE and can be solved exactly. In the following we will provide many such examples and present their explicit solutions in {\it closed form} few of which were known.

\para
In this extended introduction we present the framework within which the flow equation is derived as well as our general strategy toward its solution. The remainder of the paper consists of many examples of QFTs whose flow can be followed exactly. Section \ref{sec:2d} is dedicated to examples in two dimensions. Generalisation to higher dimensions is discussed in section \ref{sec:hd} and we close with some concluding remarks in section \ref{sec:con}.

\para

\subsection*{The $T\bar{T}$ flow equation}

Let $M$ denote a two dimensional manifold equipped with a (Euclidean) metric tensor $g_{\mu\nu}$ with $\mu,\nu=1,2$ and consider a QFT on $M$ whose dynamics is governed by the local action
\begin{equation*}
	S_{\circ}=\int_M \rmd^2x \sqrt{g}\, \cL_{\circ}(\Phi,g_{\mu\nu},\lambda) \, .
\end{equation*}
Here $\cL_{\circ}$ denotes the Lagrangian for the local fields which we have collectively denoted by $\Phi$. The coupling constants, denoted by $\lambda$, control the strength of interactions among the fields as well as with local sources.
The partition function of this theory,
\begin{equation*}
	\cZ_{\circ}[g_{\mu\nu}, \lambda] = \int [\cD\Phi]\, e^{-S_{\circ}} \,,
\end{equation*}
thus depends on the constants $\lambda$ as well as the background metric $g_{\mu\nu}$. The $T\bar T$-flow equation is the first order differential equation in a real deformation parameter $t$,
\begin{equation}
\label{qTTB}
	\left( \frac{\partial}{\partial t} + \Delta \right) \cZ_{t}=0 \,,
\end{equation}
where the functional operator $\Delta$ above is defined as
\begin{equation}
	\Delta= \lim_{\delta \to 0}\int_M \rmd^2x \frac{2}{\sqrt{g}}
	\epsilon^{\mu\nu} \epsilon^{\rho\sigma}
	\frac{\delta}{\delta g^{\mu\rho}(x+\delta)}
	\frac{\delta}{\delta g^{\nu\sigma}(x-\delta)} \,.
\end{equation}
The initial condition for \eqref{qTTB} is provided by the undeformed theory $\cZ_{t=0} = \cZ_{\circ}$.
Once the initial condition is given, then the solution is uniquely determined.

\para
For Lagrangian theories, equation (\ref{qTTB}) becomes the equation for the
action functional
\begin{equation}
\frac{\partial S}{\partial t} = (S,S)
\label{aTTB}
\end{equation}
where the pairing $(\,\cdot\,,\,\cdot\,)$ is defined via
\begin{equation*}
	(X,Y)=\lim_{\delta \to 0}\int_M \rmd^2x \frac{2}{\sqrt{g}} \epsilon^{\mu\nu} \epsilon^{\rho\sigma}
	\frac{\delta X}{\delta g^{\mu\rho}(x+\delta)}
	\frac{\delta Y}{\delta g^{\nu\sigma}(x-\delta)}\,,
\end{equation*}
for {\it local} functionals $X$ and $Y$. Equation \eqref{aTTB} is derived in \cite{Smirnov:2016lqw}, where the absence of contact terms in the $T\bar T$ composite operator is proven to follow from general assumptions\footnote{This is analogous to the absence of contact terms for the YM curvature which facilitates the derivation of Migdal's loop equations in Yang-Mills theories. Furthermore, the analogue of Polyakov's loop Laplacian for YM is $\Delta$ above (See \S 7.2 in \cite{Polyakov:1987ez}). }.
This implies that the point splitting regulator $\delta$ in the definition of the $T\bar T$ composite operator can be removed after the regularisation of the QFT and does not compete with its UV regulator.

\para
Our approach to integrating $T\bar T$-variations is concretely obtained by giving a class of solutions of  \eqref{aTTB}. The computation of the path integral for the deformed theory is a different issue
which we do not address here and we restrict our analysis to the deformation problem of the classical action.
We propose a simple integration technique for equation \eqref{aTTB} which follows from the locality of the action, the absence of space-time derivatives in $\Delta$ and covariance under diffeomorphisms ${\rm Diff}(M)$.

\para
Let us summarise our logic. The first observation is that equation \eqref{aTTB}, being first order in $t$, has a \emph{unique} solution for any initial (undeformed) local action. Locality of the operator $\Delta$ therefore suggests that we should look for a solution which can be expressed as a local functional $S(t)=\int_M \rmd^2x \sqrt{g}\, \cL(t)$ at finite $t$.
We next observe that the deformation operator $\Delta$ does not generate terms involving derivatives of the metric unless such terms are already present in the undeformed Lagrangian\footnote{In the absence of such terms therefore the deformed Lagrangian can be viewed as a function only of certain combinations of the dynamical fields and couplings. The form of these ``\emph{invariants}" is dictated by the flow equation \eqref{lTTB} as well as the explicit dependence of the undeformed Lagrangian on the metric. This, as it will be shown later, induces very strict dependences on the metric and allows the complete integration of the $T\bar T$-flow equation.}.
Therefore, the above assumptions enable us to recast \eqref{aTTB} as a local equation for the Lagrangian density ${\cL}$:
\begin{equation}
\label{lTTB}
	\partial_{t} \cL = \cL^{2} - 2\cL g^{\mu\nu} \frac{\partial \cL}{\partial g^{\mu\nu}}+ 2\varepsilon^{\mu\nu}\varepsilon^{\rho\sigma} \frac{\partial \cL}{\partial g^{\mu\rho}}\frac{\partial \cL}{\partial g^{\nu\sigma}}\,.
\end{equation}

\para
We will discuss the implementation of this method in the specific cases in the next sections. The upshot is that equation \eqref{lTTB} reduces to a partial differential equation in the deformation variable $t$ and invariants formed from the metric and the dynamical fields. We will show that in many examples the flow equation can be recast as the (extended) Burgers' equation. Since the Burgers' equation can be reduced to quadratures via the method of characteristics, we can present the explicit solution depending on the form of the initial condition.

\para
We remark that the link between the Burgers' equation and the $T\bar T$-deformed action was already observed in \cite{Cavaglia:2016oda}, where the appearance of its characteristic curve was rebuilt from the assumption of validity of the non-linear integral equation for the theory. Our approach leads directly to the Burgers' equation and in a more general setting.

\para
In the following we will first analyze the example of a single massless scalar field to familiarise the reader with our approach and to set up the notation. We then solve the case of an interacting scalar field in closed form and for an arbitrary potential before considering a general $\sigma$-model with an arbitrary target metric and $B$-field and the WZW model. We also discuss the result of a power expansion of the solution of the $T\bar T$-deformation equation in the case in which a curvature coupling is turned on and show the proliferation of higher order derivatives at higher orders in the deformation parameter. In section \ref{sec:fer} we explicitly solve the $T\bar T$-deformation of a massive Dirac fermion with quartic interaction, i.e. the massive Thirring model, and show that the solution is in this case given by a finite power series in $t$. We dedicate section \ref{sec:hd} to possible higher dimensional generalisations and close with some concluding remarks and open questions in section \ref{sec:con}.
\section{$T\bar{T}$ flows in closed form}
\label{sec:2d}

Let us first review the steps that lead from \eqref{aTTB} to the local equation \eqref{lTTB}. Consider a theory $\cT_{\circ}$ on a two dimensional manifold $M$ endowed with the Euclidean metric tensor $g_{\mu\nu}$ whose dynamics is captured by the local action $S_{\circ} = \int \rmd^{2}x \, \sqrt{g}\, \cL_{\circ}$. We are interested in finding a solution to the flow equation \eqref{aTTB} in terms of a local functional
\begin{equation}
\label{act}
	S(t) = \int \rmd^{2}x \, \sqrt{g}\, \cL(t)\,,
\end{equation}
with the initial condition $\cL(t=0) = \cL_{\circ}$. Plugging \eqref{act} into the \emph{rhs} of \eqref{aTTB} we find
\begin{equation*}
	(S,S) = \int \rmd^{2}x  \, \cO_{T\bar{T}}\,,
\end{equation*}
where the (local) $T\bar{T}$-operator is given by
\begin{equation*}
	\cO_{T\bar{T}} = \frac{1}{2} \varepsilon^{\mu\nu}\varepsilon^{\rho\sigma} T_{\mu\rho}T_{\nu\sigma}\,.
\end{equation*}
  Since the theory described by $S(t)$ is coupled to the background metric tensor $g_{\mu\nu}$, the associated energy momentum tensor can be extracted by looking at small variations of the metric,
\begin{equation*}
	T_{\mu\nu} = \frac{-2}{\sqrt{g}} \frac{\delta S(t)}{\delta\, g^{\mu\nu}} = g_{\mu\nu} \cL(t) - 2 \frac{\partial \cL(t)}{\partial g^{\mu\nu}}\,,
\end{equation*}
where the second equality holds under the condition that the undeformed Lagrangian
depends algebraically on the metric tensor. The unique solution of the $T\bar{T}$-flow equation
will, as already mentioned, enjoy the same property.
Using this expression for the energy momentum tensor enables us to recast the $T\bar{T}$-operator as
\begin{equation}
\label{TTbar}
	\cO_{T\bar{T}} = \cL^{2} - 2\cL g^{\mu\nu} \frac{\partial \cL}{\partial g^{\mu\nu}}+ 2\varepsilon^{\mu\nu}\varepsilon^{\rho\sigma} \frac{\partial \cL}{\partial g^{\mu\rho}}\frac{\partial \cL}{\partial g^{\nu\sigma}}\,.
\end{equation}
Therefore eq.\eqref{aTTB} reads
\begin{equation}
	\partial_{t} \cL = \cO_{T\bar{T}} \,,
\end{equation}
which is precisely equation \eqref{lTTB}. We study this equation in a few examples below starting with the simplest case of a free massless scalar field.

\subsection{Free massless scalar field}
As our first example we would like to find the unique solution to equation \eqref{lTTB} with the initial condition provided by the action for a free real scalar field
\begin{equation}
	S_{\circ}=\frac{1}{2}\int \rmd^{2}x \, \sqrt{g}\, g^{\mu\nu}\partial_{\mu}\phi\partial_{\nu}\phi \,.
\end{equation}
In the following we find it convenient to define the symmetric --  and metric independent -- tensor
\begin{equation*}
	X_{\mu\nu} := \partial_{\mu}\phi\partial_{\nu}\phi
\end{equation*}
whose trace we denote by $X=g^{\mu\nu} X_{\mu\nu}$. Since the initial Lagrangian is simply
\begin{equation*}
	\cL_{\circ} = \frac{1}{2}X\,,
\end{equation*}
we expect the deformed Lagrangian to depend on the fields only through $X_{\mu\nu}$. Moreover, since any diffeomorphism invariant function of $X_{\mu\nu}$ and the metric is only a function of the scalar $X$ we conclude that the deformed Lagrangian is only a function of two scalar variables $t$ and $X$, \emph{i.e.} $\cL = \cL(t,X)$. Consequently the deformation operator \eqref{TTbar} takes the simple form
\begin{equation*}
	\cO_{T\bar{T}} = \cL^{2} -2 \cL X \partial_{X} \cL \,,
\end{equation*}
yielding the flow equation
\begin{equation}
\label{fbfloweq}
	\partial_{t} \cL + (X\partial_{X}-1)\cL^{2} = 0\,.
\end{equation}
As alluded to in the introduction, this equation is simply the Burgers' equation
\begin{equation}
\label{burgers}
	\partial_{t} f(t,y) + f(t,y)\, \partial_{y} f(t,y) =0\,,
\end{equation}
with the identification
\begin{equation*}
	f(t,y) = \frac{\cL(t,X)}{\sqrt{X}}\,\quad\text{and}\quad y=\frac{-1}{\sqrt{X}}\,.
\end{equation*}
The Burgers' equation, supplemented with the boundary condition
\begin{equation*}
	f(0,y) = -\frac{1}{2y}\,,
\end{equation*}
has a unique solution which can easily be determined via the method of characteristic curves\footnote{
The Burgers' equation $\partial_{t} f(t,y) + f(t,y)\, \partial_{y} f(t,y) =0$ with boundary condition $f(0,y)=F(y)$
is equivalent to the implicit equation $f(t,y)=F\left(y-tf(t,y)\right)$.
}. The unique solution for the deformed Lagrangian is given by
\begin{equation}
\label{mb}
	\cL(t,X) = -\frac{1}{2t} + \frac{1}{2t}\sqrt{ 1 +2tX  }\,,
\end{equation}
which satisfies the boundary condition
\begin{equation*}
	\cL(0,X) = \frac{1}{2}X\,.
\end{equation*}
Note that the solution \eqref{mb} is smooth for $t\ge 0$ but can become imaginary for $t<0$. This is closely related to the fact that the spectrum of the deformed theory on a circle exhibits Hagedorn behavior for $t<0$.
The analysis above extends to more general boundary conditions which we discuss below.

\subsection{Interacting scalar field}

An immediate generalisation of the above result follows from the altered boundary condition
\begin{equation*}
	\cL(0,X) = \frac{1}{2}X+V
\end{equation*}
where $V=V(\phi)$ is an arbitrary potential so long as it is independent of the background metric. With this boundary condition corresponding to an interacting scalar the solution to \eqref{fbfloweq} is
\begin{equation}
	\cL(t,X) = -\frac{1}{2t} \frac{1-2tV}{1 - tV} + \frac{1}{2t}\sqrt{\frac{ (1 - 2tV )^{2}}{(1 - tV)^{2}} +2t\frac{X+2V}{1 - tV}  }
\end{equation}
This agrees with the expression obtained in \cite{Cavaglia:2016oda} (See also \cite{Tateo:2017igst})\footnote{After the submission of the present manuscript we became aware that the above result was presented by R.~Tateo at IGST2017. }  whose first few terms were first presented in \cite{Dubovsky:2013ira}.

\subsection{Curvature couplings}

Another generalisation of \eqref{mb} is obtained by imposing as the boundary condition a Lagrangian with curvature couplings. As an example, consider the undeformed Lagrangian
\begin{equation}
\label{cneq1}
	\cL(t=0)  = \frac{1}{2}X + \alpha_{\circ} \phi R
\end{equation}
where $R$ denotes the Ricci scalar associated with the background metric $g_{\mu\nu}$. This Lagrangian describes a theory with central charge $c=1+6Q^2$, where $\alpha_{\circ}=\sqrt{2\pi}Q$. We may think of \eqref{cneq1} as a deformation of the free theory and thus expand the solution to the flow equation in powers of $\alpha_{\circ}$,
\begin{equation*}
	\mathcal{L}=\sum_{n=0}^{N}\alpha _{\circ }^{n}\mathcal{L}^{(n)} \qquad\text{and}\qquad  T_{\mu \nu }=\sum_{n=0}^{N}\alpha _{\circ }^{n}T^{(n)}_{\mu \nu }\,,
\end{equation*}
where
\begin{equation*}
T^{(n)}_{\mu\nu} = -\frac{2}{\sqrt{g}} \frac{\delta}{\delta g^{\mu\nu}} \int\rmd^{2}x \sqrt{g} \cL^{(n)}\,.
\end{equation*}
Using this expansion we can solve the flow equation,
\begin{equation*}
\partial _{t}\mathcal{L}=\frac{1}{2}\varepsilon ^{\mu \rho }\varepsilon
^{\upsilon \sigma }T_{\mu \nu }T_{\rho \sigma }\,,
\end{equation*}
order by order in $\alpha_{\circ}$. At order $\alpha_{\circ}^{0}$ we recover \eqref{mb}, while the flow equation at order $\alpha _{\circ }$ reads
\begin{equation*}
\partial _{t}\mathcal{L}^{(1)}=\varepsilon ^{\mu \rho }\varepsilon ^{\upsilon
\sigma }T^{(0)}_{\mu \nu }T^{(1)}_{\rho \sigma }\,.
\end{equation*}
This equation can in turn be solved order by order in $t$ with the first few terms given by
\begin{equation*}
\mathcal{L}^{(1)}=\phi R-2tX\square \phi +2t^{2}X^{2}\square \phi -%
\frac{8}{3}t^{3}X^{3}\square \phi +\mathcal{O}\left( t^{3}\right)\,.
\end{equation*}
This leads us to consider the following ansatz
\begin{equation*}
\mathcal{L}^{(1)}=\phi R+f\left(tX\right) \square \phi\,.
\end{equation*}
Plugging this ansatz in the flow equation yields
\begin{equation*}
2\sqrt{1+2y}+f^{\prime }\left( y\right) y\sqrt{1+2y}+2q\left( y\right) -2=0\,,
\end{equation*}
where $y=tX$, and
\begin{equation*}
q\left( y\right) =1+\int \rmd y\,\frac{yf^{\prime}(y)}{2\sqrt{1+2y}}\,.
\end{equation*}
Solving for $f\left( y\right) $ we obtain the deformed Lagrangian
\begin{equation}
\mathcal{L}(t)=-\frac{1}{2t}+\frac{1}{2t}\sqrt{1+2tX}+\alpha _{\circ }\left[
\phi R-\log \left( 1+2tX\right) \square \phi \right] +\mathcal{O}\left(
\alpha _{\circ}^{2}\right) .
\end{equation}
As one might have expected, upon deformation, the Ricci scalar term induces higher derivative corrections with the second order derivative term $\square\phi=\nabla^\mu\partial_\mu\phi$ appearing at order $\alpha_{\circ}$. The $\alpha_{\circ}$-expansion of the deformed Lagrangian therefore takes the form of an expansion in higher derivative terms which have proved too cumbersome to determine.

\subsection{Non-linear $\sigma$-model}
\label{sec:sm}

Now that the logic is clear lets see if we can generalise the above analysis to multiple scalar fields described by the $\sigma$-model action
\begin{equation*}
	S_{\circ} =\frac{1}{2} \int \rmd^{2}x \, \sqrt{g}\, \left[ g^{\mu\nu} \partial_{\mu} \phi^{i} \partial_{\nu}\phi^{j} G_{ij}(\phi) + \varepsilon^{\mu\nu} \partial_{\mu} \phi^{i} \partial_{\nu} \phi^{j} B_{ij}(\phi) \right]\,.
\end{equation*}
As before we define a set of metric independent
tensors
\begin{equation*}
	X^{ij}_{\mu\nu} = \partial_{\mu} \phi^{i} \partial_{\nu}\phi^{j}\,,
\end{equation*}
and the
(density) scalars
\begin{equation*}
	  X^{ij} = g^{\mu\nu} X^{ij}_{\mu\nu}\quad\text{and}\quad \tilde{X}^{ij} = \sqrt{g}\varepsilon^{\mu\nu} \partial_{\mu} \phi^{i} \partial_{\nu} \phi^{j} \,.
\end{equation*}
Note that the scalar densities $\tilde{X}^{ij}$ are independent of the background metric. In fact the entire $B$-term is metric independent and therefore topological. Furthermore, topological terms are not affected by continuous, non-geometric, parameter deformations of the theory. The upshot is that the topological $B$-term is unaffected by the deformation and does not enter the analysis below.

\para
The deformed Lagrangian depends on the metric only through the worldsheet scalars $X^{ij}$ and $\tilde{X}^{ij}/\sqrt{g}$. Moreover, the latter only depends on the metric through the factor $\Omega=\sqrt{g}$. Therefore the deformed Lagrangian is expected to be a function of the deformation parameter $t$, the variables $X^{ij}$ and of $\Omega$, \emph{i.e.}
\begin{equation*}
	\cL = \cL(t,X^{\ij},\Omega)\,.
\end{equation*}
This allows us to considerably simplify the expression for the deformation operator \eqref{TTbar}
\begin{equation*}
	\cO_{T\bar{T}} = \frac{2\tilde{X}^{ik} \tilde{X}^{jl}}{\Omega^{2}}\frac{\partial \cL}{\partial X^{ij}}\frac{\partial \cL}{\partial X^{kl}} - 2\Omega\frac{\partial\cL}{\partial \Omega} X^{ij}\frac{\partial \cL}{\partial X^{ij}} + \Omega^{2} \left(\frac{\partial\cL}{\partial\Omega}\right)^{2}+ \left(1- X^{ij}\frac{\partial}{\partial X^{ij}}+\Omega\frac{\partial}{\partial \Omega}\right) \cL^{2}\,.
\end{equation*}
Note that factors of $\tilde{X}^{ij}$ in this equation should be treated as constant coefficients as they do not depend on the metric.

\para
So far we have only insisted on invariance under worldsheet diffeomorphisms. However, we expect the deformed Lagrangian to also be invariant under target space diffeomorphisms. This further constrains the form of the deformed Lagrangian such that it can only depend on the scalar $X=G_{ij}X^{ij}$, and the above equation simplifies to
\begin{equation}
\label{smTTB}
	\partial_{t} \cL =\frac{2\tilde{X}^{ij} \tilde{X}_{ij}}{\Omega^{2}} \left(\frac{\partial \cL}{\partial X}\right)^{2} - 2\Omega\frac{\partial\cL}{\partial \Omega} X\frac{\partial \cL}{\partial X} +\Omega^{2} \left(\frac{\partial\cL}{\partial\Omega}\right)^{2}+ \left(1- X\frac{\partial}{\partial X}+\Omega\frac{\partial}{\partial \Omega}\right) \cL^{2}\,.
\end{equation}
The solution to the above equation is given by
\begin{equation}
	\cL = -\frac{1}{2t} +\frac{1}{2t}\sqrt{1+2tX+2t^{2}\tilde{X}^{ij} \tilde{X}_{ij}\Omega^{-2}}\,,
\label{eee}
\end{equation}
satisfying the boundary condition $\cL(t=0)=X/2$.
The solution \eqref{eee} is valid for arbitrary target space metric, generalising the case of a flat metric which already appeared in \cite{Cavaglia:2016oda}. As was explained the $B$-term does not enter the analysis and is only introduced through $\cL_{B}(t=0)=X/2+ B_{ij}\tilde{X}^{ij}\Omega^{-1}$ resulting in the deformed action
\begin{equation}
	\cL = -\frac{1}{2t} +\frac{1}{2t}\sqrt{1+2tX+2t^{2}\tilde{X}^{ij} \tilde{X}_{ij}\Omega^{-2}}+ B_{ij}\tilde{X}^{ij}\Omega^{-1}\,.
\end{equation}

\subsection{WZW model}

The analysis of $\sigma$-models in section \ref{sec:sm} can readily be applied to Wess-Zumino-Witten (WZW) models. For simplicity we limit the discussion to the case of $SU(N)$ WZW theory described by the action
\begin{equation}
	S_{\circ}=\frac{k}{8\pi }\int_{M}\rmd^{2}x\sqrt{g}g^{\mu \nu }\Tr\left(\gamma^{-1}\partial _{\mu }\gamma \gamma ^{-1}\partial _{\nu}\gamma\right) +\frac{ik}{16\pi ^{2}}\int_{B^{3}}\left[ \Tr\gamma ^{-1}\rmd\gamma \right] ^{3}
\end{equation}
where $B^{3}$ is any three manifold whose boundary is the the worldsheet $M$. The second term
in the WZW action above is topological and thus, as explained in section \ref{sec:sm}, does not enter the flow equation and can be treated as a shift in the initial conditions.

\para
In order to use the results of the previous section we first have to express the WZW action in the $\sigma$-model variables. To this end we define the $su(N)$-valued vector field
\begin{equation*}
	A_{\mu }^{a}t^{a}=\gamma^{-1}\partial _{\mu }\gamma
\end{equation*}
where $t^{a}$ denote the generators of the $su(N)$. In analogy with the $\sigma$-model analysis of the previous section we define the scalars
\begin{equation*}
	X^{ab} = \frac{k}{4\pi }g^{\mu\nu}A_{\mu}^{a}A_{\nu}^{b}\,,
\end{equation*}
and the scalar densities
\begin{equation*}
	\tilde{X}^{ab} = \frac{k\sqrt{g}}{4\pi }\varepsilon^{\mu\nu}A_{\mu}^{a}A_{\nu}^{b}\,.
\end{equation*}
The deformed Lagrangian satisfies the same equation \eqref{smTTB} as any $\sigma$-model. The resulting deformed action is therefore given by
\begin{equation}
	S=\int_{M}\rmd^{2}x\sqrt{g}\left[-\frac{1}{2t} +\frac{1}{2t}\sqrt{1+2t X+2t^{2}\tilde{X}^{ij} \tilde{X}_{ij}\Omega^{-2}}\right] +\frac{ik}{16\pi ^{2}}\int_{B^{3}}\left[ \Tr\gamma ^{-1}\rmd\gamma \right] ^{3}\,,
\end{equation}
where the terms under the square-root are expressed in terms of the original fields $\gamma$ as
\begin{equation*}
	X = \frac{k}{4\pi} g^{\mu\nu}\Tr\left(\gamma^{-1}\partial _{\mu }\gamma \gamma ^{-1}\partial _{\nu}\gamma\right)\,,
\end{equation*}
and
\begin{equation*}
	\frac{\tilde{X}^{ij} \tilde{X}_{ij}}{\Omega^{2}} = \left(\frac{k}{4\pi}\right)^{2} \varepsilon^{\mu\nu} \varepsilon^{\rho\sigma}\Tr\left(\gamma^{-1}\partial _{\mu }\gamma \gamma ^{-1}\partial _{\rho}\gamma\right)\Tr\left(\gamma^{-1}\partial _{\nu }\gamma \gamma ^{-1}\partial _{\sigma}\gamma\right)\,.
\end{equation*}
%
%
%
%
%

\subsection{Massive Thirring model}
\label{sec:fer}

We now turn our attention to theories with fermionic fields. Consider a single Dirac fermion with the undeformed action
\begin{equation}
\label{ffaction}
	S_{\circ} = \int \rmd^{2}x \sqrt{g} \left[ \frac{i}{2}  \left(\bar{\psi} \gamma^{\mu} \nabla_{\mu} \psi - \nabla_{\mu}\bar{\psi}\gamma^{\mu}\psi \right) + V \right]\,,
\end{equation}
where the potential is given by
\begin{equation*}
	V = - m \bar{\psi}\psi +\frac{\lambda}{4} \bar{\psi}\gamma^{a}\psi\ \bar{\psi}\gamma_{a}\psi \,.
\end{equation*}
The covariant derivative acts on the fermions via
\begin{equation}
\label{scd}
	\nabla_{\mu}\psi = \partial_{\mu}\psi +\frac{i}{2} \omega_{\mu} \gamma^{3} \psi,
	\qquad
	\nabla_{\mu}\bar{\psi} = \partial_{\mu}\bar{\psi} - \frac{i}{2} \omega_{\mu} \bar{\psi}\gamma^{3},
	\qquad
	\gamma^{\mu} = e^{\mu}_{a} \gamma^{a}\,,
\end{equation}
and the spin connection (in two dimensions) satisfies
\begin{equation*}
	\omega^{ab}_{\mu} = \epsilon^{ab} \left(\frac{1}{2}\epsilon_{cd}\omega^{cd}_{\mu}\right)=\epsilon^{ab}\omega_{\mu}\,,
\end{equation*}
with $\epsilon^{12}=-\epsilon^{21}=1$. To study the $T\bar{T}$-flow of this theory we first define the $2\times 2$ matrix $X$ as follows
\begin{equation*}
	X_{ab} = \frac{i}{2}\left(\bar{\psi}\gamma_{a}\nabla_{b}\psi - \nabla_{b}\bar{\psi}\gamma_{a}\psi\right)\,.
\end{equation*}
Here $a$ and $b$ are flat indices which are raised, lowered and contracted with the flat (Euclidean) metric $\delta^{ab}$. Using \eqref{scd} one can show that
\begin{equation*}
	X_{a\mu} = e_{\mu}^{b}X_{ab} =  \frac{i}{2}\left(\bar{\psi}\gamma_{a}\nabla_{\mu}\psi - \nabla_{\mu}\bar{\psi}\gamma_{a}\psi\right)
	= \frac{i}{2}\left(\bar{\psi}\gamma_{a}\partial_{\mu}\psi - \partial_{\mu}\bar{\psi}\gamma_{a}\psi\right) \,.
\end{equation*}
which is manifestly independent of the metric. Since the undeformed Lagrangian \eqref{ffaction} is simply $\cL_{\circ} = \Tr X + V = e^{a\mu}X_{a\mu}+V$, we can work out the energy momentum tensor of the undeformed theory
\begin{equation}
\label{ff0em}
	T^{(0)}_{ab} = \frac{2}{\sqrt{g}} e_{a}^{\mu}e_{b}^{\nu} \frac{\delta S^{(0)}}{\delta g^{\mu\nu}}
	= 2 e_{a}^{\mu}e_{b}^{\nu} \frac{\partial e^{\lambda}_{c}}{\partial g^{\mu\nu}}X_{c\lambda} - \delta_{ab} \cL^{(0)}
	= X_{(ab)} - \delta_{ab} \left(\Tr X+V\right) \,.
\end{equation}
It is clear that the deformed Lagrangian is constructed solely from $X$ and $V$ and since these only contain the fermionic fields $\psi$, $\bar{\psi}$ and their first derivatives we conclude that the deformed Lagrangian can only contain products of up to order $X^{4}$, $X^{2}V$ and $V^{2}$ as all higher powers vanish identically. We therefore expect the $t$-expansion of the deformed Lagrangian to terminate. Consequently we expand the Lagrangian and the energy momentum tensor of the deformed theory as
\begin{equation}
\label{ffLt}
	\cL = \sum_{n=0}^{N} t^{n} \cL^{(n)} \qquad\text{and}\qquad  T_{\mu\nu} =  \sum_{n=0}^{N} t^{n} T^{(n)}_{\mu\nu}\,,
\end{equation}
where
\begin{equation*}
	T^{(n)}_{\mu\nu} = -\frac{2}{\sqrt{g}} \frac{\delta}{\delta g^{\mu\nu}} \int\rmd^{2}x \sqrt{g} \cL^{(n)}\,.
\end{equation*}
Employing this expansion we can solve the flow equation\footnote{Here we have used the identity $g^{\mu\nu}g^{\rho\sigma} - g^{\rho\nu}g^{\mu\sigma} =  \varepsilon^{\mu\rho}\varepsilon^{\nu\sigma}$ which holds true in two dimensions. }
\begin{equation}
\label{floweq}
	\partial_{t} \cL = \frac{1}{2} \varepsilon^{\mu\rho}\varepsilon^{\nu\sigma} T_{\mu\nu}T_{\rho\sigma} = \frac{1}{2} \left(g^{\mu\nu}T_{\mu\nu}\right)^{2} - \frac{1}{2} T^{\mu\nu}T_{\mu\nu} \,,
\end{equation}
order by order. Using \eqref{ffLt}, our flow equation \eqref{floweq} at order $t^{n-1}$ reads
\begin{equation*}
	\cL^{(n)} = \frac{1}{2n}(g^{\mu\nu}g^{\rho\sigma} - g^{\mu\sigma}g^{\rho\nu} ) \sum_{i+j=n-1}(2-\delta_{ij})T^{(i)}_{\mu\nu} T^{(j)}_{\rho\sigma} \,.
\end{equation*}
Since $\cL^{(n)}$ only depends on the metric through $X$ we can apply the chain rule to obtain
\begin{equation*}
	2 e_{a}^{\mu}e_{b}^{\nu} \frac{\partial \cL}{\partial g^{\mu\nu}} = X_{c(a}\delta_{b)d}  \frac{\partial \cL}{\partial X_{cd}}  \,.
\end{equation*}
Furthermore, as we will see below $\cL^{(n)}$ only depends on the symmetric part of $X$ which we will denote by $\tilde{X}_{ab}=X_{(ab)}$.
We can now solve \eqref{floweq} order by order starting from the undeformed energy momentum tensor \eqref{ff0em}. At order $t^{0}$ we have
\begin{equation*}
	\cL^{(1)} =  \frac{1}{2} \left(\Tr X \right)^{2} - \frac{1}{2}\Tr(\tilde{X}^{2} ) +  V^{2} + V \Tr X \,,
\end{equation*}
from which we can evaluate $T^{(1)}_{ab}$ as follows
\begin{equation*}
	T^{(1)}_{ab} = \Tr \tilde{X} \tilde{X}_{ab} - \left(\tilde{X}X\right)_{(ab)} + V \tilde{X}_{ab} -2\delta_{ab} \cL^{(1)} \,.
\end{equation*}
Next we analyze the flow equation \eqref{floweq} at order $t^{1}$ which, after dividing by $2t$, reads
\begin{equation*}
	\cL^{(2)} = \frac{1}{4}\Tr \tilde{X}^{3} -\frac{3}{8} \Tr \tilde{X}\Tr\tilde{X}^{2} + \frac{1}{8} \left(\Tr\tilde{X}\right)^{3} +\frac{V}{4} \left((\Tr \tilde{X})^{2} - \Tr\tilde{X}^{2} \right) \,.
\end{equation*}
The corresponding contribution to the energy momentum tensor is
\begin{equation*}
\begin{aligned}
	T^{(2)}_{ab} =&\, \frac{3}{4} (\tilde{X}^{2}X )_{(ab)}-\frac{3}{4}(\Tr \tilde{X} ) (\tilde{X}X)_{(ab)}-\frac{3}{8}\left(\Tr\tilde{X}^{2}-(\Tr\tilde{X})^{2}\right) \tilde{X}_{ab}
	\\
	&+\frac{V}{2} \left( (\tilde{X}X)_{(ab)} - \Tr\tilde{X}\, \tilde{X}_{ab} \right) -\delta_{ab} \cL^{(2)} \,.
\end{aligned}
\end{equation*}
The final term in the $t$-expansion of the Lagrangian is determined by equating the terms at order $t^{2}$ in \eqref{floweq}. At this order we find
\begin{equation*}
	\cL^{(3)} = -\frac{1}{6} \Tr\tilde{X}^{4} + \frac{1}{12} (\Tr\tilde{X}^{2})^{2} +\frac{1}{4} \Tr\tilde{X}\,\Tr\tilde{X}^{3}-\frac{5}{24}(\Tr\tilde{X})^{2}\Tr\tilde{X}^{2}+\frac{1}{24} (\Tr\tilde{X})^{4} \,.
\end{equation*}
Note that the higher order terms in the $t$-expansion of the flow equation \eqref{floweq} vanish identically thanks to the Grassmann nature of the fermionic fields from which $X$ is built. The final form of the $T\bar{T}$-deformed Lagrangian is therefore
\begin{equation}
\label{TTff}
\begin{aligned}
	\cL =& \Tr\tilde{X} +V + \frac{t}{2}\left((\Tr\tilde{X})^{2} - \Tr\tilde{X}^{2} + 2V(V+\Tr\tilde{X})\right)
	\\
	& +\frac{t^{2}}{2}\left(\Tr\tilde{X}^{3}-\frac{3}{2}\Tr\tilde{X}\,\Tr\tilde{X}^{2}+\frac{1}{2}(\Tr\tilde{X})^{3}+V(\Tr\tilde{X})^{2}-V\Tr\tilde{X}^{2}\right)
	\\
	& -\frac{t^{3}}{3}\left(2 \Tr\tilde{X}^{4} -(\Tr\tilde{X}^{2})^{2} -3 \Tr\tilde{X}\,\Tr\tilde{X}^{3} + \frac{5}{2}(\Tr\tilde{X})^{2}\Tr\tilde{X}^{2}-\frac{1}{2} (\Tr\tilde{X})^{4}\right)\,,
\end{aligned}
\end{equation}
where $\tilde{X}$ is given by
\begin{equation*}
	\tilde{X}_{ab} = \frac{i}{2} \left(\bar{\psi}\gamma_{(a}\nabla_{b)}\psi - \nabla_{(a}\bar{\psi}\gamma_{b)}\psi\right)\,.
\end{equation*}
By expanding \eqref{TTff} and using Fierz identities one can in fact show that the expression for the Lagrangian drastically simplifies to
\begin{equation*}
	\cL = \cL^{(0)} + \frac{t}{4}\left((\Tr\tilde{X})^{2} - \Tr\tilde{X}^{2} + 2m^{2} (\bar{\psi}\psi)^{2}-2m\bar{\psi}\psi\Tr\tilde{X}\right)-\frac{t^{2}}{8}m\bar{\psi}\psi\left((\Tr\tilde{X})^{2}-\Tr\tilde{X}^{2}\right)\,.
\end{equation*}
The explicit form of the $T\bar{T}$-deformed Lagrangian in (flat) complex coordinates is
\begin{equation}
\begin{aligned}
	\cL(t) =\ & i\left(\bar{\psi}_{-} \overleftrightarrow{\partial_{z}} \psi_{-} - \bar{\psi}_{+} \overleftrightarrow{\partial_{\bar{z}}} \psi_{+}\right) - m\left(\bar{\psi}_{-}\psi_{+} - \bar{\psi}_{+}\psi_{-}\right)
	+(\lambda -m^{2}t) \bar{\psi}_{+}\bar{\psi}_{-} \psi_{+}\psi_{-}
	\\
	& -\frac{imt}{2}\left[\bar{\psi}_{+}\bar{\psi}_{-} \left(\psi_{-}\partial_{z}\psi_{-}-\psi_{+}\partial_{\bar{z}}\psi_{+}\right) +\psi_{+}\psi_{-}\left(\bar{\psi}_{-}\partial_{z}\bar{\psi}_{-} - \bar{\psi}_{+}\partial_{\bar{z}}\bar{\psi}_{+}\right)\right]
	\\
	& + \frac{t}{4}\left[\left(\bar{\psi}_{+}\overleftrightarrow{\partial_{\bar{z}}}\psi_{+}\right) \left(\bar{\psi}_{-}\overleftrightarrow{\partial_{z}}\psi_{-}\right)
	+\bar{\psi}_{-} \partial_{z} \bar{\psi}_{-} \psi_{-}\partial_{z}\psi_{-} +\bar{\psi}_{+}\partial_{\bar{z}}\bar{\psi}_{+} \psi_{+}\partial_{\bar{z}}\psi_{+}
	-2\left(\bar{\psi}_{+}\overleftrightarrow{\partial_{z}}\psi_{+}\right)\left(\bar{\psi}_{-}\overleftrightarrow{\partial_{\bar{z}}}\psi_{-}\right)\right]
	\\
	& -\frac{mt^{2}}{8} \bar{\psi}_{+}\bar{\psi}_{-} \psi_{+}\psi_{-} \left(\partial_{z}\bar{\psi}_{-}\partial_{\bar{z}}\psi_{+} - \partial_{\bar{z}}\bar{\psi}_{+}\partial_{z}\psi_{-}
	-2\partial_{\bar{z}}\bar{\psi}_{-}\partial_{z}\psi_{+} + 2 \partial_{z}\bar{\psi}_{+}\partial_{\bar{z}}\psi_{-}\right)\,.
\end{aligned}
\end{equation}

\para
We stress that the expansion in the deformation parameter terminates. This is akin to the observation made in \cite{Guica:2017lia} for a Lorentz-breaking irrelevant deformation analogous to $T\bar{T}$. Consequently we anticipate that for the deformed theory to receive an infinite series of corrections, as is the case for the Goldstino \cite{Zamolodchikov:1991vx}, we need to turn on an infinite tower of irrelevant deformations.

\section{Generalisation to higher dimensions}
\label{sec:hd}

Let us consider higher dimensional generalisations of the $T\bar T$-deformations. Such a generalisation was recently proposed by J.~Cardy \cite{Cardy:2018sdv} in the form of $|\det T|^{1/\alpha}$ with $\alpha=D-1$ in $D$ dimensions. We will treat this generalisation, for more general values of the parameter $\alpha$, in some detail later in the section.

\para
Let us remark in passing that there is another possible generalisation of the $T\bar T$-deformation which remains quadratic in the energy momentum tensor. Starting in two dimensions we first use the identity $\epsilon^{\mu\nu}\epsilon^{\rho\sigma}=g^{\mu\rho}g^{\nu\sigma}-g^{\nu\rho}g^{\mu\sigma}$. This suggests the following $D$-dimensional generalisation of the flow equation
\begin{equation*}
 \partial_t S= \frac{1}{2}\int \rmd^{D}x\,\sqrt{g}
 \left[
 \left(g^{\mu\rho}g^{\nu\sigma}-g^{\nu\rho}g^{\mu\sigma}\right)
  \left(\frac{-2}{\sqrt{g}}\frac{\delta S}{\delta g^{\mu\rho}}\right)
 \left(\frac{-2}{\sqrt{g}}\frac{\delta S}{\delta g^{\nu\sigma}}\right)
 \right].
\end{equation*}
For a single scalar field -- without conformal couplings -- the flow equation for the Lagrangian takes the form
\begin{equation*}
\partial_t {\mathcal L}=(D-1)\left[(D/2){\mathcal L}^2-2X\partial_X{\mathcal L}{\mathcal L}\right]\,,
\end{equation*}
where $X=\partial_\mu\phi g^{\mu\nu}\partial_\nu\phi$, which reduces to the Burgers' equation. Although we do not treat this case further in this work, let us note that this could have a more natural AdS dual interpretation compared to the $(\det T)^{1/\alpha}$. Whether either of these generalisations can be defined at the quantum level remains an open question. We remark that the scaling solution, i.e. with ${\mathcal L}_{t=0}=\frac{1}{2}X$, of this equation is given by solving the algebraic equation $\left(1+\frac{D(D-1)}{2}t{\mathcal L}\right)^{4-D}{\mathcal L}^D=\left(X/2\right)^D$.
Therefore, the free massless scalar in four dimensions is a fixed point of the flow and one needs to turn on a potential (or a conformal coupling) to have a nontrivial evolution.

\para
We now turn our attention to the flow instigated by an operator of the form $(-\det T)^{1/\alpha}$  in $D$-dimensions resulting in the flow equation
 \begin{equation}
 \partial_t S= \frac{1}{\alpha-D}\int \rmd^{D}x\,\sqrt{g}
 \left[\frac{-1}{D!}
 \epsilon^{\mu_1\ldots\mu_D}\epsilon^{\nu_1\ldots\nu_D}
 \left(\frac{-2}{\sqrt{g}}\frac{\delta S}{\delta g^{\mu_1\nu_1}}\right)
 \ldots
 \left(\frac{-2}{\sqrt{g}}\frac{\delta S}{\delta g^{\mu_D\nu_D}}\right)
 \right]^{1/\alpha}
\label{detT}
\end{equation}
 where $\alpha$ is a real parameter\footnote{The case of \cite{Cardy:2018sdv} is $\alpha=D-1$, but we keep this parameter free to different examples.}. For this to be an irrelevant deformation for CFTs we take $0<\alpha<D$ and we further assume $\alpha$ to be an integer.
Let us integrate the above equation in the case of a scalar field $\phi$ by reducing it to a partial differential equation. Once more we define $X=\partial_\mu\phi g^{\mu\nu}\partial_\nu\phi$ and write the solution as
\begin{equation*}
	S=\int \rmd^{D}x \sqrt{g} {\mathcal L}(X,t)
\end{equation*}
with the initial condition
\begin{equation*}
	{\mathcal L}(X,0) = \frac{1}{2} X + V
\end{equation*}
for a generic local potential $V=V(\phi)$. Since the deformed Lagrangian only depends on the background metric through $X$ the expression for the associated energy momentum tensor simplifies to
\begin{equation*}
	\frac{-2}{\sqrt{g}}\frac{\delta S}{\delta g^{\mu\nu}}=g_{\mu\nu}{\mathcal L}-2 \partial_X{\mathcal L} \partial_\mu\phi\partial_\nu\phi\,.
\end{equation*}
Using the above expression, equation \eqref{detT} simplifies to
\begin{equation*}
	\partial_t{\mathcal L} = \frac{1}{\alpha-D} \left[{-\mathcal L}^D+2{\mathcal L}^{D-1}(X\partial_X){\mathcal L}\right]^{1/\alpha}.
\end{equation*}
This can be further simplified by considering the redefinition
\begin{equation*}
	Y=(-1)^{\alpha} X^{\frac{\alpha-D}{2}} \quad {\rm and }\quad {\mathcal L}=\sqrt{X} f^{\frac{1}{D-\alpha}}\,,
\end{equation*}
yielding
\begin{equation}
\label{a-burgers}
	(\partial_t f)^\alpha+ f^\alpha\partial_Y f=0\,,
\end{equation}
which reduces to the Burgers' equation for $\alpha=1$. The relevant initial condition is obtained by inverting the above redefinitions in $f=\left(X^{-1/2}{\mathcal L}\right)^{D-\alpha}$ and $X=\left[(-1)^{\alpha} Y\right]^{\frac{2}{\alpha-D}}$. The initial condition therefore takes the form
\begin{equation*}
	f(0,Y)=(-1)^{\alpha} Y\left\{\frac{1}{2}\left[(-1)^{\alpha} Y\right]^{\frac{2}{\alpha-D}}+V \right\}^{D-\alpha}\,.
\end{equation*}
Below we solve \eqref{a-burgers} with this initial condition in a few cases.

\para
The solution of the Burgers' equation, that is (\ref{a-burgers}) with $\alpha=1$, is given by solving the implicit equation,
\begin{equation*}
	f(t,Y)=f\left(0,Y-tf(t,Y)\right)\,,
\end{equation*}
which for our initial condition reads
\begin{equation}
\label{general}
	\frac{f}{tf-Y}=\left[\frac{1}{2}\left(tf-Y\right)^{\frac{2}{1-D}}+V \right]^{D-1}\,.
\end{equation}

\para
Exact solutions of equation (\ref{general}) can be obtained explicitly for low values of the dimension $D$ and for a generic potential. The solution drastically simplifies in the case of the massless free scalar, \emph{i.e.}  $V=0$, and reads as
\begin{equation}
\label{mlfD}
	f(t,Y)=\frac{1}{2t}\left(Y+\sqrt{Y^2+\frac{t}{2^{D-3}}}\,\right)\,.
\end{equation}
This results in the deformed Lagrangian
\begin{equation}
\label{masslessD}
	{\mathcal L}_{D,1}(t,X)=\left\{\frac{1}{2t}\left[\sqrt{1+4t(X/2)^{D-1}}-1\right]\right\}^{1/(D-1)}\,.
\end{equation}

\para
Another interesting case which can be simplified is the scaling solution for the free massless scalar field and arbitrary value of $\alpha$. In this case the differential equation,
\begin{equation}
\label{a-burgers'}
	(\partial_t f)^{\alpha}+ f^{\alpha}\partial_Y f=0\,,
\end{equation}
is accompanied by the simple boundary condition $f(0,Y)=\frac{(-1)^\alpha}{2^{D-\alpha}}\frac{1}{Y}$. We can further reduce eq. (\ref{a-burgers'}) by using scaling symmetry. This enables us to set
\begin{equation*}
	f(t,Y)=(-\alpha/2)^\alpha t^{-\frac{\alpha}{2}} K\left(Yt^{-\frac{\alpha}{2}}\right)\,,
\end{equation*}
and reduces the PDE above to an ODE in the variable $Z=Yt^{-\frac{\alpha}{2}}$:
\begin{equation*}
	0=\left(1+Z\frac{K^{\prime}}{K}\right)^\alpha+K^{\prime} \,.
\end{equation*}
This equation should have a solution such that $K\sim \frac{1}{kZ}$ for large positive $Z$ and with $k=2^{D-2\alpha}(\alpha)^{\alpha}$. After some thought one can see that such a solution indeed exists. Of course, the case $\alpha=1$ reproduces the solution \eqref{mlfD} for the Burgers' equation. The explicit solution for the first few small integer values of $\alpha$ can be computed by reducing to quadratures. For example, at $\alpha=2$ we find
\begin{equation*}
	K=\frac{1}{c^2Z+c}\, \quad {\rm with} \quad c^2=2^{D-2},
\end{equation*}
where the two possible signs of $c$ are related by the $t\to-t$ symmetry of the equation. The corresponding Lagrangian (with $c=2^{(D-2)/2}$) is
\begin{equation}
\label{alpha2free}
	\cL_{D,2}=\frac{X}{2}\frac{1}{\left[1+t(X/2)^{\frac{D-2}{2}}\right]^{\frac{1}{D-2}}}\,.
\end{equation}
In particular, for $D=3$, the above Lagrangian is the result of integrating the $(\det T)^{\frac{1}{D-1}}$ deformation proposed in \cite{Cardy:2018sdv} for a free scalar field theory in three dimensions which takes the simple form
\begin{equation*}
	\cL_{3,2}=\frac{X}{2}\frac{1}{1+ t\sqrt{X/2}}\,.
\end{equation*}

\section{Conclusions and Open questions}
\label{sec:con}

In this paper we have presented a general approach to $T\bar T$-deformations of quantum field theories in two dimensions, as well as some generalisations to higher dimensions, and demonstrated its effectiveness in a number of cases, notably non-linear $\sigma$-models and the massive Thirring model.

\para
There are many other theories of interest in two dimensions to which our approach can be applied, notably Yang-Mills theories and gauged linear $\sigma$-models as well as their supersymmetric counterparts. Moreover, the method presented here can be extended to flows instigated by analogs of the $T\bar T$-operator involving higher spin currents proposed in \cite{Smirnov:2016lqw} or symmetry breaking currents such as the one discussed in \cite{Guica:2017lia}.

\para
One of the most pressing questions left unanswered is the issue of extending the exact integration method to theories whose undeformed action includes curvature couplings, such as the conformal coupling term for scalars in higher dimensions. As was demonstrated in a simple example in two dimensions, curvature couplings can be treated order by order but a closed form solution eludes us.

\para
Another crucial issue yet to be addressed with regards to the higher dimensional generalisations of the $T\bar{T}$ operator is to analyze the existence or absence of contact terms, along the lines of \cite{Zamolodchikov:2004ce}, for the composite operator $\left(\det T\right)^{1/\alpha}$ in $D>2$ and for different values of $\alpha$. It is reasonable to expect that the quantum corrected operator has additional terms, especially in the light of the expected form of the deformation equation for the partition function proposed in \cite{Cardy:2018sdv}. Needless to say the holographic interpretation of $T\bar{T}$-like deformations in higher dimensions is also of great interest and needs to be addressed.

\para
\para
{\bf Acknowledgements}: We would like to thank J.~Elias Mir\'o for many fruitful discussions specially pertaining to section \ref{sec:fer}.
This research is supported by the PRIN project ``Non-perturbative Aspects Of Gauge Theories And Strings''  and by INFN Iniziativa Specifica ST\&FI.


%
%
%
%
%
%
%
%
%
%
%
%
%
%
%
%
%
%
%
%
%
%
%
%

\bibliography{refs}
\bibliographystyle{utphys}

\end{document}